\documentclass[twoside,twocolumn,english]{revtex4}
\usepackage[T1]{fontenc}
\usepackage[latin1]{inputenc}
\usepackage{graphicx}

\makeatletter

\usepackage{graphicx}

\usepackage{babel}
\makeatother
\begin{document}

\title{Accelerating Universes from Short-Range Interactions.}

\author{Alberto Díez-Tejedor%
\footnote{wtbditea@lg.ehu.es%
} and Alexander Feinstein%
\footnote{a.feinstein@ehu.es%
}}

\address{Dpto. de Física Teórica, Universidad del País Vasco, Apdo. 644, 48080,
Bilbao, Spain.}

\begin{abstract}
We show that short-range interactions between the fundamental particles
in the universe can drive a period of accelerated expansion. This
description fits the early universe. In the present day universe,
if one postulates short-range interactions or a sort of \char`\"{}shielded
gravity\char`\"{}, the picture may repeat. 
\end{abstract}
\maketitle
There is no doubt that the inflationary paradigm has played a central
role during the last several decades in our understanding of the early
universe \cite{Guth}. While most of the authors rarely question the
initial acceleration itself, the underlying cause of the acceleration
remains rather obscure. It is often assumed that the initial accelerated
expansion is driven by a sort of a self interacting scalar field \cite{Belinsky},
the so-called \emph{inflaton}. Much work has been done during the
last twenty years studying the possible form of the potential for
the scalar field which best fits within the cosmological model \cite{Potentials}.
Recently, other proposals have been put forward where the inflaton
takes the form of a non-canonical scalar field, also known as K-field
\cite{K-field}. Needless to say that the inflaton is rather an effective
field which most probably does not correspond to any fundamental particle,
but could be looked at as an effective description of an underlying
physical theory \cite{Vega}.

A related issue is the state of the matter at the very high densities
which prevail in the early universe. Presumably, the fundamental theories
such as Superstrings, M-theory, etc. \cite{string} might guide one
as to what symmetries and laws to expect when the densities, velocities
and energies approach those present near the Big Bang. Certain clues
about the behavior of the ultradense matter can be also obtained in
the framework of Quark-Gluon Plasma theory \cite{QGP}, which could
be handy to model the description of the primordial matter in the
early universe. This area of research is now under intensive theoretical and experimental study, mainly at
RHIC and at CERN, with the surprising new observation that in  this extreme state the
matter seems to behave almost as a non-viscous perfect fluid 
\cite{RHIC}. At this stage, however, there is little one can say
with a certain degree of rigor about the properties of matter nearly
the initial singularity. 

This brings us to the following question: Is there a way to obtain
inflationary solutions in a highly dense universe within a so-to-say
{}``conventional physics''? Physics, which one might label as fairly
realistic and, to some extent, generic. The answer to this question
is the main purpose of this Letter, and follows to be affirmative.

We will be considering a universe filled with an ultradense matter
and modeling the interaction between the particles by a short-range
attractive force. For computational purposes we will be using the
Yukawa type potentials, yet, our discussion is generic and applies
to \textit{any} attractive short-range interaction, as is explained
below. Moreover, our findings hold through as well in the case of
the diluted matter, but in this case, one would need to postulate
some unknown short-range interaction, or assume that the Newtonian
gravity in the expanding universe is somehow {}``shielded''. We
will comment on these issues before closing. Our main result is, that
describing the matter as a fluid of interacting particles via short-range
attractive forces, the cosmological models undergo a phase of accelerated
expansion.

The introduction of short-range forces between the particles leads
to the following equation of state for the matter: \begin{equation}
\rho=m_{0}n-An^{2},\quad p=-An^{2},\label{eq:1}\end{equation}
 or equivalently, in a $p(\rho)$-form, \[
p=\rho-\frac{m_{0}^{2}}{2A}\left[1-\sqrt{1-\frac{4A}{m_{0}^{2}}\rho}\right].\]
 Here $A$ is a term depending on the interaction, $m_{0}$ is the
rest mass of the fundamental particle and $n$ is the particle number
density. The conventions and units we use are $c=8\pi G=\hbar=1$
and the metric has a signature $\left(-,+,+,+\right)$.

This, somewhat unusual in cosmology equation of state, is the one
which could drive the accelerated expansion in the early universe.
We see that if one neglects the interaction between particles ($A=0$),
or consider a very dilute matter ($n\rightarrow0$), the equation
of state (\ref{eq:1}) reduces to the standard dust equation $\rho=m_{0}n$,
$p=0$. Here we  impose also the positivity of the energy density ($n<m_{0}/A$).

To obtain the above equation of state (see for example \cite{Bludman}),
we consider a system of $N$ identical interacting particles placed
at points $\mathbf{l}_{i}$ ($i=1,2,...,N$). The interaction between
the particles is modeled by a potential $V\left(l_{ij}\right)$, where
we have defined $l_{ij}\equiv\left|\mathbf{l}{}_{i}-\mathbf{l}{}_{j}\right|$.
We further neglect the temperature effects by assuming that the particle masses and the strength of the interaction are much larger than the effects of the temperature,  and write the energy of
the system as\[
U=m_{0}N+\frac{1}{2}\sum_{i=1}^{N}\sum_{j\neq i}V(l_{ij}).\]
 Dividing by the volume and assuming that the system is homogeneous
and isotropic, at least at the scale of the interaction, one readily
obtains the energy density as a function of $n$ as given in the equation
(\ref{eq:1}). The pressure, then, is computed using the thermodynamic
relation $p=n\left(\partial\rho/\partial n\right)-\rho$. The interaction
term $A$ can be evaluated passing as usual to the continuous limit
and replacing the sum by the integral: \begin{equation}
A=-2\pi\int_{0}^{\infty}V(l)l^{2}dl.\label{eq:a}\end{equation}
 Note, that the sign of the parameter $A$ defines the sign of the
pressure, and therefore, in order to obtain an accelerated expansion,
the interaction should be attractive.

A typical way of modeling a short range interaction is via the
Yukawa potential $V(l)=-g^{2}e^{-\mu l}/l$. Here $g$ is the coupling
constant of the theory and $\mu$ the mass of the boson mediating
the force, whose Compton length $l_{0}=1/\mu$ defines the range of
the interaction. The integral of the equation (\ref{eq:a}), therefore,
is unaffected by the upper limit at infinity. Performing the integral,
we obtain: \[
A=\frac{2\pi g^{2}}{\mu^{2}}.\]

It is important to say that the convergence of the above integral
basically defines as to whether the interaction is short- or long-range,
and whether the system of particles may be treated by conventional
thermo- and hydro-dynamics. If the potential energy due to the interaction
between the particles falls faster than $r^{-3}$, the force is short-range,
the above integral converges and the usual thermodynamics apply. Such
is the case for the Yukawa potential or, for example,  the  Van der Waals forces, etc. If, however, the potential energy
does not fall as fast as $r^{-3}$, as for example in the case of
Newtonian gravity one deals with long-range
forces, and the dynamics of the systems governed by those may be quite
involved \cite{Padma}. There is little doubt, however, that the forces
between the fundamental particles in the early universe are short-range,
therefore providing a broad motivation for our further analysis.

We have found it useful to express the pressure, using the thermodynamic
relations, in terms of the enthalpy per particle $h$: \begin{equation}
p(h)=-\frac{1}{4A}\left(h-m_{0}\right)^{2},\label{eq:2}\end{equation}
 where the enthalpy is given by \begin{equation}
h=\frac{\partial\rho}{\partial n}=m_{0}-2An.\label{eq:4}\end{equation}

We further assume that the ultradense matter is described by an isentropic
irrotational perfect fluid, and to proceed, we give the action from
which the equations of the fluid motion are derived \cite{Schutz,nosotros}:
\[
S=\int d^{4}x\sqrt{-g}\left\{ p\left(\left|V\right|\right)-\left(\frac{\partial p}{\partial h}\right)\left[\left|V\right|-\frac{V^{\mu}\phi_{,\mu}}{\left|V\right|}\right]\right\} .\]
 Here the current $V^{\mu}$, usually known as the Taub current, is
given by $V^{\mu}=hu^{\mu}$ ($\left|V\right|=h$), where $u^{\mu}$,
the 4-velocity of the fluid, verifies $u_{\mu}u^{\mu}=-1$. The dynamical
variables are $g^{\mu\nu}$, $V^{\mu}$ and $\phi$, and the following
equations of motion result: \begin{equation}
u_{\mu}=-h^{-1}\phi_{,\mu},\label{eq:din1}\end{equation}
\begin{equation}
\left(nu^{\mu}\right)_{;\mu}=0,\label{eq:din2}\end{equation}
 with the stress-energy tensor given by \begin{equation}
T^{\mu\nu}=\frac{2}{\sqrt{-g}}\frac{\delta S}{\delta g_{\mu\nu}}=\left(\frac{\partial p}{\partial h}\right)hu^{\mu}u^{\nu}+pg^{\mu\nu}.\label{eq:momento}\end{equation}
 The usual thermodynamic relation between the density and the pressure,
$\rho=nh-p$ (with $n=\left(\partial p/\partial h\right)$), allows
one to cast the energy-momentum tensor into the standard perfect fluid
form: \[
T^{\mu\nu}=\left(\rho+p\right)u^{\mu}u^{\nu}+pg^{\mu\nu}.\]
 The equation (\ref{eq:din1}) is the Clebsch decomposition of the
4-velocity for an irrotational flow, whereas the equation (\ref{eq:din2})
is the conservation law for the particle number. Equation (\ref{eq:din1})
shows that the scalar field $\phi$ plays the role of a velocity potential.

The identity $u_{\mu}u^{\mu}=-1$ and the equation (\ref{eq:din1})
lead to the following expression for the enthalpy in terms of the
derivatives of the velocity potential: \begin{equation}
h=\sqrt{-g^{\mu\nu}\phi_{,\mu}\phi_{,\nu}}.\label{eq:entalpia}\end{equation}
 Taking into account the Clebsch decomposition of the 4-velocity (\ref{eq:din1}),
and introducing the new variable $X=-\frac{1}{2}g^{\mu\nu}\phi_{,\mu}\phi_{,\nu}=h^{2}/2$,
we obtain the on-shell expression for the action: \begin{equation}
S_{on-shell}=\int d^{4}x\sqrt{-g}F(X),\label{eq:action}\end{equation}
 where in the final expression we have defined\begin{equation}
p(h)=p\left(\sqrt{2X}\right)\equiv F(X).\label{eq:def2}\end{equation}

Expression (\ref{eq:action}) gives the action for an irrotational
perfect fluid. This functional takes the form of a non-canonical scalar
field action and is analogous to the one often used in K-essence cosmology
(known as purely kinetic K-essence) \cite{K-field,nosotros,Armendariz,Scherrer}.
The Lagrangian density is obtained from the equation of state which
relates the pressure with the enthalpy (\ref{eq:def2}), and the scalar
field is the velocity potential of the fluid, which plays the role
of the inflaton field. For completeness, the density and particle
number in terms of the variable $X$ are given: \begin{equation}
\rho=2XF'(X)-F(X),\quad n=\pm\sqrt{2X}F'(X),\label{eq:presion}\end{equation}
 where $F'(X)$ denotes the derivative of the function with respect
to its variable.

The above formalism together with the equation of state in the form
(\ref{eq:2}) and the assumption that one deals with an irrotational
perfect fluid, gives the following matter Lagrangian: \begin{equation}
F_{\pm}(X)=-aX\pm b\sqrt{X}-\frac{b^{2}}{4a},\label{eq:3a}\end{equation}
 where the parameters $a$ and $b$ are given by: \[
a=\frac{1}{2A},\quad b=\frac{m_{0}}{\sqrt{2}A}.\]
 The $\pm$ sign in the equations appears due to the definition of
the enthalpy ($h=\pm\sqrt{2X}$) and separates the matter into an
ordinary one, in the case of positive enthalpy, and the so-called
\emph{phantom matter} ($\rho+p<0$) \cite{phantom}, in the case where
the enthalpy is negative. It is interesting that in the case of the
perfect fluid, the phantom and the ordinary matter have a simple physical
separation depending whether the enthalpy per particle is negative
or positive respectively.

The behavior of the function $F(X)$ is depicted in the figure 1.
Note, that the function $F(X)$ is defined by the physical parameters
of the system: $m_{0}$ and $A$ and is not assumed \textit{ad hock}
as is usually done in the context of the K-essence cosmology. The
positivity of the energy and of the number of particles imposes $X\leq X_{s}$
for both the ordinary and the phantom branch. The condition of the
accelerated expansion $\rho+3p\leq0$ is always verified in the phantom
case, whereas in the case of the ordinary matter one needs $X\leq X_{r}$
for inflation.

\begin{figure}
\begin{center}\includegraphics[%
  scale=0.38]{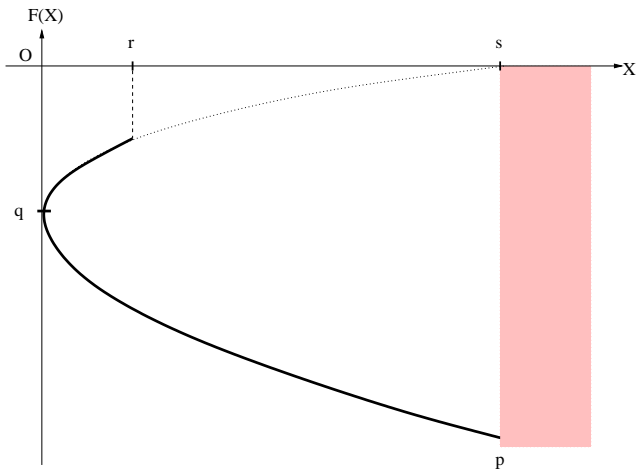}\end{center}

\caption{The figure shows the shape of the function $F$ in terms of the kinetic
term $X$. The upper branch, above the point $q$, represents the
standard matter, whereas the lower branch dives into the phantom domain.
The solid line represents the values of $X$ for which inflation occurs.
The values of the different points are: $X_{s}=X_{p}=b^{2}/4a^{2}$,
$X_{r}=b^{2}/16a^{2}$ and $F(0)=-b^{2}/4a$.}
\end{figure}

We now consider an isotropic and homogeneous universe, modeled for
simplicity by the spatially flat Friedmann-Robertson-Walker metric:
\begin{equation}
ds^{2}=-dt^{2}+R^{2}(t)\left[dr^{2}+r^{2}\left(d\theta^{2}+\sin^{2}\theta d\varphi^{2}\right)\right].\end{equation}
 $R(t)$ is the scale factor and ($t,r,\theta$,$\varphi$) the comoving
coordinates. The physical distances in such a universe are given by
$l=R(t)r$. The dynamical equation for the field $X$ can be obtained
either from the particle conservation equation (\ref{eq:din2}) or
by the direct variation of the on-shell action and results in: \begin{equation}
\left[F'(X)+2XF''(X)\right]\dot{X}+6HF'(X)X=0.\label{eq:dinamic}\end{equation}
 Here, $H=\dot{R}/R$ is the Hubble parameter. Solving this expression
for $\dot{X}$, we obtain: \begin{equation}
\dot{X}=-\frac{6HF'(X)X}{F'(X)+2XF''(X)}.\label{eq:dotX}\end{equation}
 The equation (\ref{eq:dotX}) basically determines the behavior of
the cosmological model \cite{Scherrer}. We assume we deal with an
expanding universe by fixing the sign of the Hubble parameter to be
positive. Since the field $X$ is positive by definition, the sign
of $\dot{X}$ is determined by the signs of $F'(X)$ and $F'(X)+2XF''(X)$.
We can check that $F'(X)+2XF''(X)=-a$ for both branches, whereas
$F'(X)$ is positive on the positive branch of our picture, and negative
on the negative one. This means, that whatever the initial value of
$X$ is, one always finishes with the dust behavior, i.e. the pressure
vanishes. If the initial value of the field $X$ falls into the phantom
branch, $X$ decreases to zero, passes to the standard matter branch
and then increases, till finally the model becomes pure dust near
$X_{s}$. The interesting point is, however, that the short-range
interaction between the particles leads to a period of accelerated
expansion naturally finished at $X_{r}$, which is an exit point towards
the dust-like universe (grateful exit).

The picture that emerges, is therefore, as follows. The short-range
attractive forces in the early universe introduce a measure of negative
pressure (tension) and could be responsible for the early universe
inflation. The inflation stops naturally at the exit point defined
by the parameters of the theory.

One should not expect from this simple model, as it stands, a reasonable amount of inflation. Indeed, our
estimates do not seem to produce  anything close to 60-70 e-foldings one would like to have for an effective inflation. The point is that
the model is too simple to acomodate the interplay between the scales of the particle horizon and the interaction range. Moreover, in the case of the  very early universe there will be instants where the horizon size would be even smaller than the range of the interaction. Then, the interaction term $A$ will be time dependent, due to the fact that the upper limit 
of the integral in the equation (\ref{eq:a}) will be bounded by the size of the  time-dependent particle horizon. This, introduces interesting physics and enhances the inflation, on one hand, but complicates the equations of the dynamical evolution on the other. 
We hope to be able to report the results of the numerical studies of these equations in the near future.

It would be extremely interesting if a similar mechanism could explain
as well the present day acceleration \cite{Acc.}. We see here two
possible alternatives. First, would be to postulate the existence
of fundamental particles which dominate the universe and interact
via short-range forces \cite{Fischbach}, in the sense that the expression
(\ref{eq:a}) converges. We will not speculate about this possibility
here, however, it should be pointed out that interesting models have
been recently proposed in which the late time acceleration of the
universe is obtained with the use of a \emph{Van der Waals} equation
of state \cite{Capozziello}, and, as is known from statistical mechanics,
this kind of equation of state appears when one takes into account the
interaction between the particles. 

Another possibility, which looks
more appealing to us and is connected again to the interplay between the horizon size and the range of the interaction, is to consider the Newtonian gravity between
the galaxies, or clusters, as the dominant contribution to the matter
density and the pressure in the present day universe. One can think of the Newtonian gravity as the interaction  between the fundamental particles in the universe. ``Averaging'' on a scale comparable with the cluster scale
one can consider this term as an effective part of the energy-momentum tensor which drives the expansion.
Now, the Newtonian
gravity is a long-range force.
Nevertheless, in an expanding past singular universe, one may apply
a natural \emph{cut off} to the integral (\ref{eq:a}) to evaluate
the interaction term $A$. Since no interaction (including gravity) acts beyond the particle
horizon, the latter may serve as the interaction range scale for the
gravity. Effectively, then, the gravity would become shielded by the horizon in an infinite past-singular expanding universe and could produce
the necessary negative pressure to accelerate the universe. We leave,
however, this  and the  related problem of the time-dependent interaction term for future report.
\begin{acknowledgments}
A.D.T. work is supported by the Basque Government predoctoral fellowship
BFI03.134. This work is supported by the Spanish Science Ministry
Grant 1/MCYT 00172.310-15787/2004, and the University of the Basque
Country Grant 9/UPV00172.310-14456/2002. 
\end{acknowledgments}


\begin{thebibliography}{10}
\bibitem{Guth}A.H. Guth, Phys. Rev. D \textbf{23}, 357 (1981); A. Linde, {}``Particle
Physics and Inflationary Cosmology'', Harwood, 1990; A.R. Liddle
and D.H. Lyth, {}``Cosmological Inflation and Large-Scale Structure'',
Cambridge University Press, 2000. 
\bibitem{Belinsky}V.A. Belinsky, L.P. Grishchuk, I.M. Khalatnikov and Ya.B Zeldovich,
Phys. Lett. B \textbf{155}, 232 (1985). 
\bibitem{Potentials}J.E. Lidsey, A.R. Liddle, E.W. Kolb, E.J. Copeland, T. Barreiro and
M. Abney, Rev. Mod. Phys. \textbf{69}, 373 (1997). 
\bibitem{K-field}C. Armendariz-Picon, T. Damour and V. Mukhanov, Phys. Lett. B \textbf{458},
209 (1999). 
\bibitem{Vega}D. Cirigliano, H.J. de Vega and N.G. Sanchez, Phys. Rev. D \textbf{71},
103518 (2005); D. Cirigliano, H.J. de Vega and N.G. Sanchez, arXiv
astro-ph/0507595.
\bibitem{string}J.E. Lidsey, D. Wands and E.J. Copeland, Phys. Rep. \textbf{337},
343 (2000); M. Gasperini and G. Veneziano, Phys. Rep. \textbf{373},
1 (2003). 
\bibitem{QGP}J. Harris and B. Müller, Ann. Rev. Nucl. Part. Sci. \textbf{46}, 71
(1996). 
\bibitem{RHIC}Results from the first 3 years at RHIC, www.bnl.gov/bnlweb/prubaf/pr/docs/Hunting-the-QGP.pdf
(to appear in Nucl. Phys. A); E. Shuryak, Prog. Part. Nucl. Phys.
\textbf{53}, 273 (2004).
\bibitem{Bludman}S.A. Bludman and M.A. Ruderman, Phys. Rev. \textbf{170}, 1176 (1968);
Ya.B. Zeldovich, JETP \textbf{14}, 1143 (1962); H.Dehnen and H. Hönl,
Astrophys. Space Sci. \textbf{33}, 49 (1975). 
\bibitem{Padma}Th. Dauxois, S. Ruffo, E. Arrimondo and M. Wilkens, {}``Dynamics
and Thermodynamics of Systems with Long Range Interaction'', Lecture
Notes in Physics \textbf{602}, 1 (2002); T. Padmanabhan, Phys. Rep.
\textbf{188}, 285 (1990). 
\bibitem{Schutz}B.F. Schutz, Phys. Rev. D \textbf{2}, 2762 (1970); B.F. Schutz and
R. Sorkin, Ann. Phys. \textbf{107}, 1 (1977); J.D. Brown, Class. Quant.
Grav. \textbf{10}, 1579 (1993). 
\bibitem{nosotros}A. Diez-Tejedor and A. Feinstein, Int. Jour. Mod. Phys. D \textbf{14}, 1561  (2005), arXiv gr-qc/0501101 
\bibitem{Armendariz}C. Armendariz-Picon, V. Mukhanov and P.J. Steinhardt, Phys. Rev. Lett.
\textbf{85}, 4438 (2000); T. Chiba, T. Okabe and M. Yamaguchi, Phys.
Rev. D \textbf{62}, 023511 (2000). 
\bibitem{Scherrer}L.P. Chimento, Phys. Rev. D \textbf{69}, 123517 (2004); R.J. Scherrer,
Phys. Rev. Lett. \textbf{93}, 011301 (2004) 
\bibitem{phantom}R.R. Caldwell, Phys. Lett. B \textbf{545}, 23 (2002); S.M. Carroll,
M. Hoffman and M. Trodden, Phys. Rev. D \textbf{68}, 023509 (2003). 
\bibitem{Acc.}A.G. Riess \emph{et al}, Astrom. J. \textbf{116}, 1009 (1998); S.
Perlmutter \emph{et al}, Astrophys. J. \textbf{517}, 565 (1999). 
\bibitem{Fischbach}E. Fischbach, D. Sudarsky, A. Szafer and C. Talmadge, Phys. Rev. Lett.
\textbf{56}, 3 (1986).
\bibitem{Capozziello}S. Capozziello, S. De Martino and M. Falanga, Phys. Lett. A \textbf{299},
494 (2002). \end{thebibliography}
\end{document}